\documentclass[11pt,a4]{article}
\usepackage{times,fancyhdr}
\usepackage[dvips]{graphicx}
\usepackage{amsmath}
\usepackage{amssymb}
\usepackage{array}
\usepackage[small,bf]{caption}
\usepackage{color}
\usepackage{fancyhdr}
\usepackage{fancyvrb}
\usepackage{hyperref}
\usepackage{longtable}
\usepackage{setspace}
\usepackage{framed}
\usepackage{titlesec}
\usepackage{titling}
\usepackage{xcolor}
\usepackage{authblk}

\fancyheadoffset[R]{0pt}
\fancyfootoffset[R]{0pt}
\definecolor{orange}{rgb}{1.0,0.5,0.0}
\definecolor{aqgr}  {rgb}{0.0,1.0,0.6} 
\definecolor{viol}  {rgb}{0.8,0.6,0.8}
\definecolor{figdr} {rgb}{1.0,1.0,1.0} 
\definecolor{colne} {rgb}{1.0,0.0,1.0} 
\definecolor{coldr} {rgb}{1.0,0.8,0.0} 
\definecolor{colop} {rgb}{0.5,1.0,1.0} 
\definecolor{colok} {rgb}{0.7,1.0,0.7} 
\definecolor{chgcol}{rgb}{0.7,0.0,0.0}

\newcolumntype{C}[1]{>{
   \centering\let\newline\\\arraybackslash\hspace{0pt}}m{#1}}
\usepackage{array}
\newcolumntype{P}[1]{>{\centering\arraybackslash}p{#1}}
\newcommand\parop[1]{\colorbox{colop}{\textbf{#1}}}

\newcommand\neudef[1]{\textcolor{black}{\textit{#1}}}
\def\afhead{0}

\title{\vspace{0.0cm}\bfseries{\textsc{ 
   Evolvable Soma Theory of Ageing: \\ 
   Insights from Computer Simulations
}}}

\author{}  
\vspace{-1.0cm} 
\date{}
\begin{document}

\pretitle{%
\begin{center}\LARGE
\vskip -2.2cm
\rule{\textwidth}{2.0pt}
\par
\vskip 0.5cm
}
\posttitle{
\par
\rule{\textwidth}{2.0pt}
\end{center}
\vskip -0.0cm
}
\maketitle

\vspace*{-2.5cm}
\begin{center}
\begin{tabular}{P{5.0cm} P{2.0cm} P{5.0cm}}
\textbf{Alessandro Fontana} & & 
\textbf{Marios Kyriazis} \\
National Gerontology Centre, Cyprus & & 
National Gerontology Centre, Cyprus \\
\texttt{fontalex00@gmail.com} & & 
\texttt{drmarios@live.it}
\end{tabular}
\end{center}
   
\vspace*{0.0cm}
\begin{abstract}
\if\afhead2 {\parop{xxxx}} \fi
Biological evolution continuously refines the design of species, resulting in highly optimised organisms over hundreds of millennia. Intuitively, we expect that random changes---evolution's primary mechanism---are more likely to be harmful than beneficial, leading to widespread detrimental effects in evolving species. The Evolvable Soma Theory of Ageing (ESTA) suggests that ageing is the cumulative result of these harmful effects, which predominantly cause bodily damage, while a few may lead to beneficial adaptations that evolution can exploit. While the disposable soma theory views aging as a consequence of limited evolutionary pressure, ESTA posits that ageing is essentially evolution in action. In this study, we gather evidence supporting this theory through computer simulations. We conduct experiments using a platform where genes are linked to onset values that determine when they are expressed. Three scenarios are tested: one with single-point fitness evaluation, constant mutation rate and fixed gene onsets; one with single-point fitness evaluation, onset-dependent mutation rate and fixed gene onsets; and one with spread fitness evaluation, onset-dependent mutation rate and evolvable gene onsets. The last scenario, which embodies the evolvable soma hypothesis, demonstrates superior performance in both algorithmic efficiency and biological plausibility compared to the others.
\end{abstract}

\section{Introduction}

Numerous theories have been put forward to interpret ageing. \textit{Stochastic theories} suggest that ageing results from damage caused by environmental factors or byproducts of cellular metabolism (such as free radicals), similar to the wear and tear of mechanical devices. \textit{Programmed theories} propose that ageing is directed by specific instructions encoded in DNA, which can be somewhat influenced by environmental conditions. \textit{Epigenetic ageing theories} \cite{Yang23epinfo} argue that changes in gene expression patterns, rather than alterations to the DNA sequence itself, are responsible for age-related changes in cellular function and tissue degeneration.

If the biological nature of ageing remains a mystery, even more perplexing is the intricate relationship between ageing and evolution, the force that shapes biological systems. The idea that ageing is nonadaptive is well-supported in the field \cite{Vijg16essence}. Ageing would persist in evolution because the pressure on an individual diminishes with age, making ageing a byproduct of limited investment in cellular repair and maintenance in favour of early-life evolutionary benefits, a concept behind the \textit{disposable soma theory} \cite{Kirkwood77,Kyriazis17}. 

Another line of research suggests that ageing evolved, just like any other trait, because it is beneficial for the individuals, or for the evolutionary process itself. Weismann proposed that death evolved to remove older individuals, promoting generational turnover \cite{Weismann91}, a concept aligned with group selection. Similarly, \cite{Goldsmith08} argued that limiting lifespan prevents older individuals from dominating the gene pool, enhancing evolutionary adaptability.

This paper is concerned with a new theory of ageing \cite{Fontana24a}, based on the assumption that development and ageing are part of the same process, driven by instructions encoded in the genome. The core of this theory is that ageing is largely driven by the cumulative effects of evolutionary ``experiments'' aimed at refining a species' design. These experiments, taking place primarily in the post-reproductive period (consistent with the principle of terminal addition), are nearly always unsuccessful due to their pseudorandom nature, resulting in bodily harm that ultimately leads to death. However, a small fraction of these experiments yield positive outcomes, providing valuable insights that guide evolutionary progress. This concept, referred to as the \neudef{Evolvable Soma Theory of Aging (ESTA)}, posits that ageing is, in essence, evolution in action.


In this study, we investigate the relationship between evolution and ageing using computer simulations, an approach with a growing presence in the field. In \cite{Dimopoulos17}, the authors perform a comparative analysis between the biological ageing paradigm and its implementation in evolutionary algorithms. They highlight a key difference in the timing of the ageing process: in nature, ageing occurs as a continuous, uncoordinated process across individuals, whereas in the algorithmic setting, it appears as a step-wise, synchronised event among individuals. \cite{Giaimo22} presents a dynamic model of life history evolution with no fixed assumptions on selection patterns. The model reveals that ageing can stably evolve, and shows a decline in selection force with age. This generalises classic ageing theory, showing that age-related selection decline emerges dynamically, not by assumption. 

In a recent study \cite{Szilágyi23}, evolution is simulated within a two-dimensional lattice arena, where agents proliferate based on fecundity and ageing, each modelled as a multilocus trait with binary alleles. The authors demonstrate that adaptive ageing can indeed evolve, a process driven by kin selection. This outcome is maintained as long as selection remains consistently directional and the rate of environmental change does not exceed the individuals' lifespans. Another study \cite{Roget24} investigates the mechanisms linking fertility timing and ageing, demonstrating in simulations that these traits are selected and stabilised over evolutionary history. This stabilisation helps explain why ageing populations tend to be more evolvable and ultimately more successful, suggesting that ageing may function as an adaptive force in evolution.

Our simulations will use the \neudef{Epigenetic Tracking (ET)} model, which enables the devo-evolution of complex three-dimensional structures in computer simulations. This model has provided insights into a range of biological phenomena \cite{Fontana12b,Fontana23a}, including ageing \cite{Fontana14,Fontana24a}. A defining feature of ET is its unified integration of development and ageing, where both processes are governed by genetically encoded events tied to onset values that evolution can alter, thereby delaying or advancing their occurrence. The findings from this model are applicable not only to ET but also to any evo-devo framework in which developmental processes are directed by programmed, time-dependent events.

\section{The ET model of development}
\label{sect:ETmodel}

\begin{figure*}[t] \begin{center} \hspace*{-0.0cm}
\includegraphics[width=15.0cm]{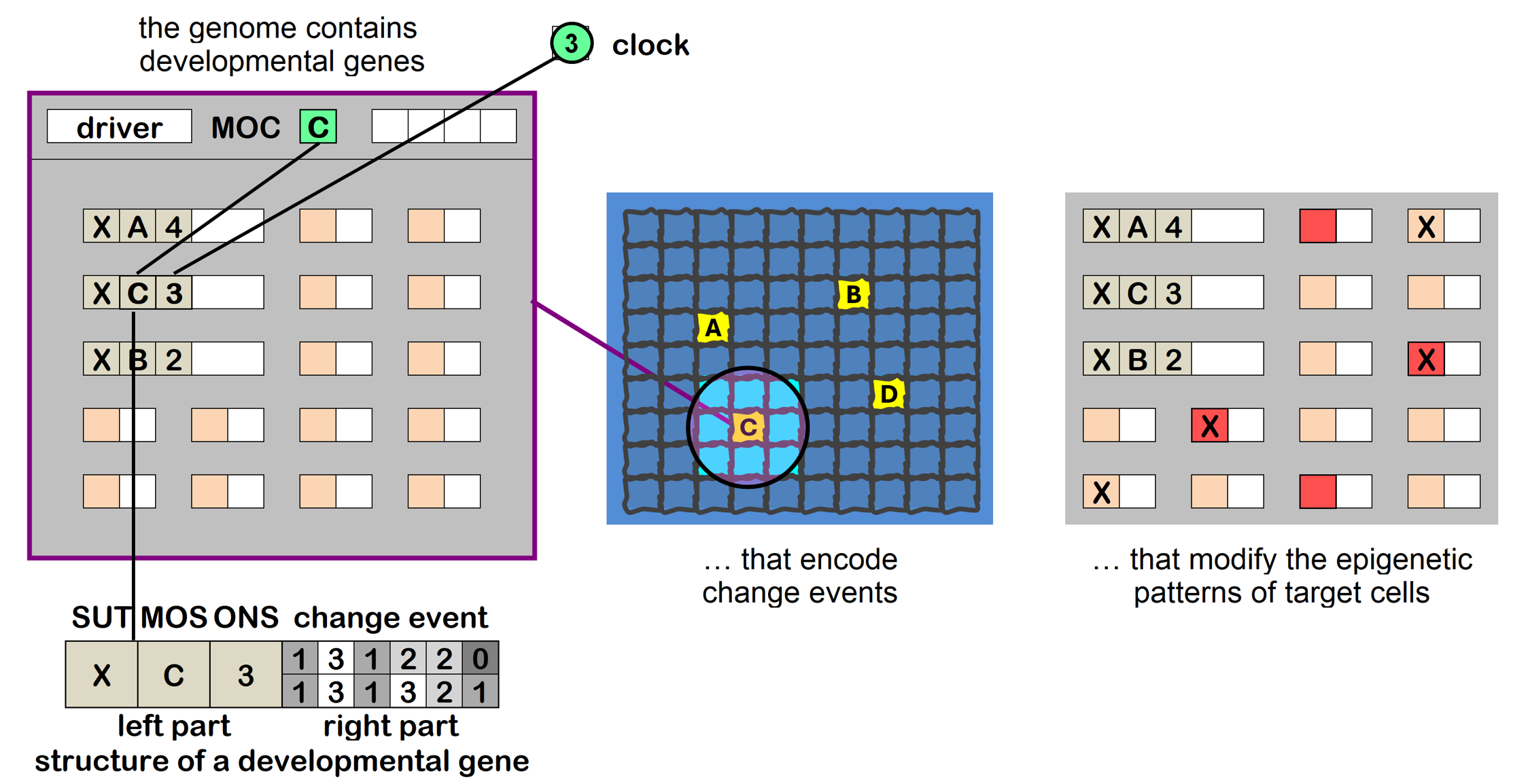}
\caption{
The genome contains developmental genes, encoding change events, which in turn shape the epigenetic landscape of target cells. The figure illustrates a change event of type differentiation. In driver cell C (left panel, top) a developmental gene is activated (middle panel). The code in gene right part (left panel, bottom) specifies the change set (middle panel) and the epigenetic changes to be applied to the initial epigenetic pattern (right panel), resulting in some genes being enabled while others are disabled.}
\label{events}
\end{center} \end{figure*}

In this model, phenotypes are represented by cellular structures composed of cube-shaped cells deployed on a grid. These cells can be categorised into two groups: \neudef{normal cells}, which constitute the majority of the structure, and \neudef{driver cells}, which are considerably fewer in quantity and uniformly dispersed throughout the structure's volume. Normal cells could be likened to basic soldiers, while driver cells might be envisioned as captains due to their ability to command normal cells. The outcome of these commands is the orchestration of \neudef{change events}, encompassing the generation, elimination, and specification of significant numbers of cells. Events are synchronised through a global clock shared by all cells.

Cells are equipped with a genome, comprising two gene categories: \neudef{developmental genes}, responsible for regulating development, and \neudef{other genes}, which perform various functions beyond the scope of this study. Apart from the genome, which remains consistent across all cells, there are also elements that exhibit variability between cells, reflecting the differentiation process inherent in development. These elements encompass the \neudef{driver mark}, distinguishing between driver and normal cells, and the \neudef{master organisation code (MOC)}, serving as an abstraction for all master regulatory elements within the cell. The biological analogous of the variable elements may include elements from the epigenome, as well as from the transcriptome.

Developmental genes can be likened to ``macro'' sets of genes that are regulated together. The left part of a developmental gene (lower left panel of Fig.~\ref{events}) is composed of three elements: i) the \neudef{structural unavailability tag (SUT)} that determines if the gene is available for activation or structurally inactive; ii) the \neudef{master organisation sequence (MOS)} that can align with the MOC; iii) the \neudef{onset (ONS)} that can match with the clock. MOC and MOS sequences are symbolised by letters, ONS and clock are shown as numbers. At each developmental step, identified by a distinct clock value, these parameters are compared for each developmental gene and driver cell. When the MOS matches the MOC and the onset matches the clock, the gene becomes active and the gene's right part is executed.

Both the MOC and the clock are responsible for initiating the activation of different segments of the genome during development, in a spatially and temporally precise manner. In biological contexts, the MOC might correspond to cell master regulatory elements, capable of initiating cascades of gene activations that dictate cellular behaviour. The clock could be implemented as a diffusible molecule, emanating from a central origin and propagated to all cells. SUTs could correspond to epigenetic elements (e.g., histone modifications or methylation marks) that determine the accessibility or structural inactivity of genes.

The right part of a developmental gene encodes a \neudef{change event}. Three types of events are foreseen. \neudef{Differentiation change events} (Fig.~\ref{events}) prompt the activated driver cell to switch the availability marks of some genes in a set of cells called \neudef{change set}), which can be so small as to involve only the driver itself. These events may correspond to writing, erasing and reading operations within the ``histone code'' framework \cite{Jenuwein01}. \neudef{Proliferation change events} lead to the proliferation of the activated driver cell (or induce the proliferation of other cells). The outcome of \neudef{apoptosis change events}, which imitate the phenomenon of biological apoptosis, is the elimination of cells within the change set from the grid.

While most normal cells generated during a proliferation event become terminally differentiated cells, some other cells revert to driver cells. The newly formed driver cells derived from normal cells acquire a new and distinct MOC value to permit differentiation. Should a developmental gene exist in the genome with a MOS corresponding to the MOC of a new driver cell, this cell can serve as the focal point for another event in a subsequent step, propelling the developmental progression. Thus, development foresees the alternation of two phases: a phase A, where driver cells produce new normal cells and induce their differentiation, and a phase B, where new driver cells emerge from normal cells. Continuously repeated, this cycle forms the core of development.

\begin{figure*}[t] \begin{center} \hspace*{-0.0cm}
\includegraphics[width=14.0cm]{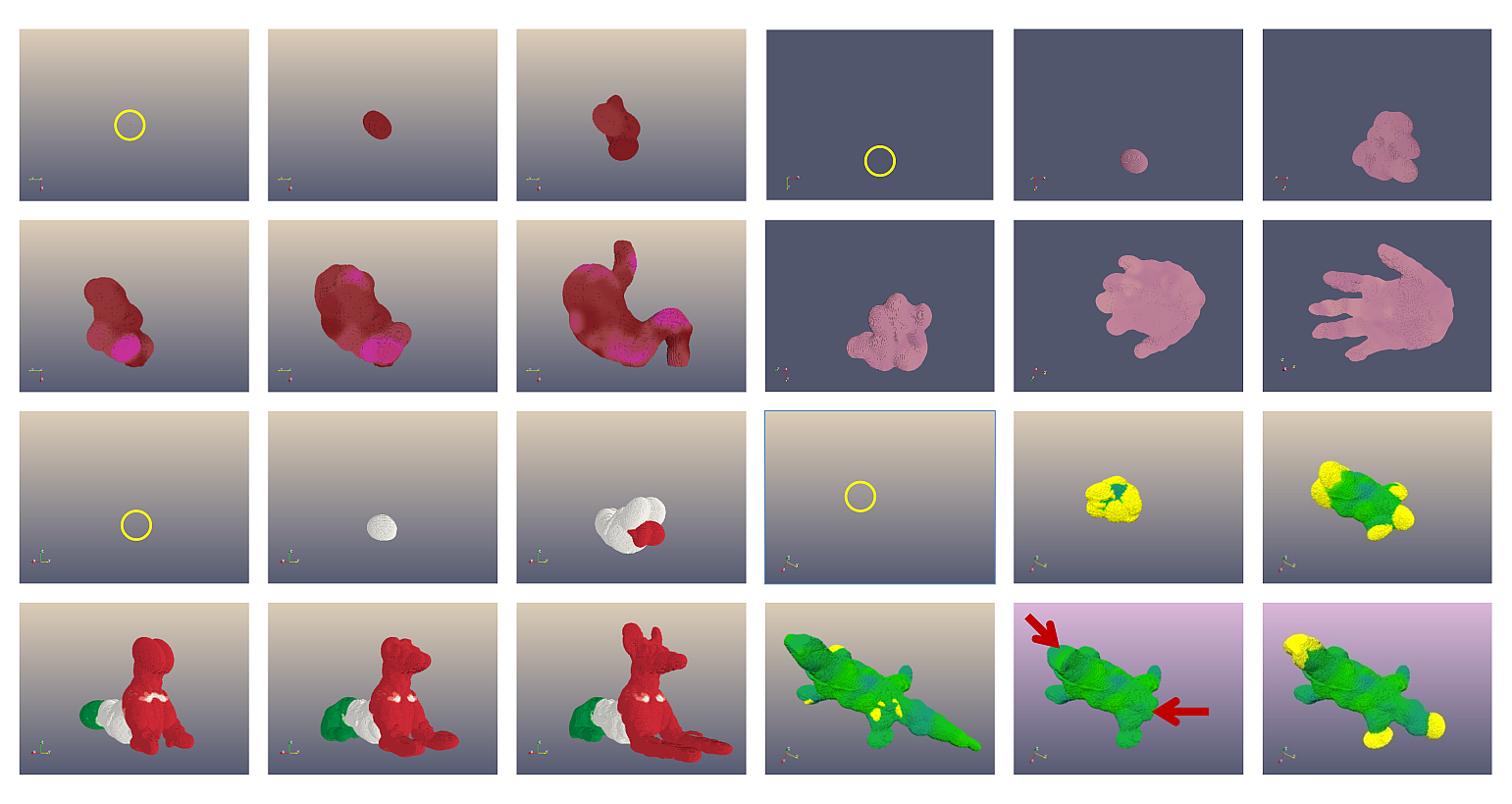}
\caption{
Examples of 3D structures generated by combining the ET model with a genetic algorithm demonstrate the model's capacity to devo-evolve complex shapes.
}
\label{shapes}
\end{center} \end{figure*}

The development model described can be integrated with a \textit{genetic algorithm} that emulates biological evolution. In this algorithm, a population of individuals, each encoded in an artificial genome, undergoes evolution over multiple generations. During each generation, all individuals progress from the zygote stage to the final phenotype, whose proximity to a predefined target serves as fitness measure. Based on these values, individuals' genomes are selected and subject to random crossover and mutations, generating a new population. This iterative cycle gives origin to a evo-devo process, that in computer simulations has demonstrated its effectiveness in producing structures of unparalleled complexity (Fig.~\ref{shapes}), encompassing both shape, regeneration capabilities and function \cite{Fontana10b}.

\section{The Evolvable Soma Theory of Ageing}

\subsection{Ageing as a program driven by time-dependent events}

In the ET model, an individual progresses through N steps of development, ultimately leading to the assessment of their fitness, which is then utilised for selection in the genetic algorithm. While in most experiments fitness was assessed at the final step (N), an alternative scenario entails allowing the global clock to continue ticking, observing events unfolding in steps N+1, N+2, ..., subsequent to fitness evaluation. This differentiation between the periods preceding and following fitness evaluation (corresponding to biologal reproduction) mirrors the biological phases of development (e.g., up to 25 years of age in humans) and ageing (from 25 years of age onward).

By the end of development, numerous driver cells are present in the individual's body. Some of these cells have been activated during development, while others (the majority in previous simulations) have not. These dormant driver cells may contain developmental genes bound to activate during the ageing phase, when, by definition, they do not impact the fitness value. Consequently, their genetic optimisation has not occurred through evolution, resulting in associated events exhibiting a ``pseudorandom'' nature, i.e. seemingly random despite being encoded in the genome, subject to genetic control, and fundamentally deterministic. This pseudorandomness implies that these events are more likely to have detrimental effects on the individual than beneficial ones.

The developmental process can be compared to a robotic painter applying brushstrokes whose characteristics---such as timing, shape, colour, and texture---are encoded in the genome. The process begins with an empty canvas and gradually unfolds to create a complete work of art. Most evolutionary theories of ageing implicitly assume that the painter stops once development concludes, leaving other processes (e.g., colour diffusion or degradation) to blur and degrade the painting over time. In contrast, the ET model suggests that the painter remains active from conception until death. However, as evolutionary pressure declines, the painter's brushstrokes, initially optimised and precise, become increasingly random, leading to progressive deterioration
 
The evolutionary pressure on a developmental gene can be represented by the absolute value of the fitness gradient with respect to the parameters influencing that gene: $\lvert \nabla F(w_i) \rvert$. The notion that a gene activated only post-reproduction experiences no selective pressure is overly simplistic. In nature, post-reproductive events, such as the quality of parental care, can significantly influence an individual's reproductive success by enhancing offspring survival, a concept encompassed by kin selection \cite{West07}. A more nuanced perspective suggests that the influence of an event on fitness---and thus the pressure on the corresponding gene---gradually lessens with the age at which the event occurs (as determined by the gene's onset value) rather than abruptly ceasing after reproduction.

Therefore, ageing is interpreted as a \textit{continuation of development}, driven by non-optimised genes activated in specific driver cells after reproduction. There is empirical evidence of an age-related decline in the functionality of adult stem cells \cite{Liu11}. Our hypothesis is that this functional decline is determined by epigenetic changes orchestrated in biological driver cells by change events associated to onset values, bound to occur at precise moments. The effects of such change events would include both manifestations of normal ``ageing'' and symptoms of ageing-associated diseases, such as type II diabetes and Alzheimer's disease.
 
\subsection{Ageing as evolution in action}

The model presented thus far is consistent with theories which views ageing as stemming from a lack of evolutionary pressure. To unlock a more profound biological meaning for ageing, we must acknowledge the formidable challenge faced by evolution: refining the design of biological species. This iterative process has persisted for hundreds of millennia, resulting for most species in a highly optimised biological blueprint. In such a scenario, the impact of random changes (the method employed by evolution) is much more likely to be harmful than advantageous for the affected individuals, potentially lowering their reproductive fitness. Hence, the effects of various detrimental changes should manifest in all individuals of evolving species. 

One crucial question arises: where do these effects manifest? The streamlined code that drives development serves as evidence of past evolution. Yet, what about present evolution? There appears to be a \textit{cause without an effect}. On the other hand, there is a phenomenon---ageing---that appears to be driven by the accumulation of random-like, deleterious events, although the molecular details are still unclear, as well as the interplay between environmental and genetic factors. In this case we have an \textit{effect without a cause}. 

Hence, there is an opportunity to catch two birds with one stone, suggesting that evolution is the cause of ageing and ageing is the effect of evolution. More specifically, our proposition is that ageing is (mostly) caused by the cumulative effect of all the ``experiments'' performed by evolution to improve a species' design. These experiments are almost always unsuccessful, as expected given their pseudorandom-like nature, cause harm to the body and ultimately lead to death. On the other hand, a small minority of experiments have positive outcome, offering valuable insight into the direction evolution should pursue. This does not rule out the possibility that other factors, such as environmental damage, may also play a role in the ageing process.

As outlined in the first section, epigenetic ageing theories suggest that age-related cellular changes and tissue degeneration are primarily driven by shifts in gene expression patterns, rather than direct alterations to DNA sequences \cite{Yang23epinfo}. Specifically, epigenetic patterns in cells appear to lose coherence with age, exhibiting a more random-like behaviour; this gradual loss of distinction between cell types contrasts sharply with the clarity of these distinctions before reproduction. Meanwhile, evolution is thought to use random DNA changes as a mechanism for introducing innovations that are then tested over time. Thus, randomness emerges as a crucial link between evolution and ageing, supporting our hypothesis.

This proposal can be viewed as the complementary perspective to the disposable soma theory, which posits that ageing is the result of an increasingly absent evolution. According to ESTA ageing, essentially, \textit{is} evolution: once an individual has fulfilled its reproductive duties, its body is not disposed of (and wasted), but used for experimentation by evolution. This is done initially with caution, as evolution is ``aware'' that the individual still has parental tasks to fulfill, and increasingly more aggressively as the offspring becomes less dependent on parental care. According to this hypothesis, ageing can be viewed as nothing less than \textit{evolution in action}. 

This raises the question: why aren't innovations integrated directly into the early developmental phase, where their impact on fitness would be greatest? We propose that evolution favours experimentation during later life stages, where ageing serves as a buffer, allowing for new changes to be introduced with a reduced fitness impact and a reduced risk. This hypothesis is consistent with concepts like ``terminal addition'' \cite{Gould77} and ``peramorphosis'' \cite{Alberch79}, which suggest that evolutionary innovations tend to appear toward the end of development.

However, evolution can also affect earlier developmental stages. Consider a scenario where a new change event arises with effects only in old age. Its fitness impact will be minimal but still positive; if beneficial, it gradually spreads in the population and is incorporated into future generations. Once the change proves ``safe'' in the ageing period, evolution may adjust the onset, shifting the event to earlier ageing steps, where the effect on fitness is stronger. With further validation, the change can eventually reach the developmental period. This process enables efficient evolution, allowing selection of advantageous traits with minimal disruption to existing designs. 



\begin{figure*}[t]
\hspace*{-0.0cm}
\begin{minipage}[c]{6.0cm} 
\centering
\small
\begin{tabular}{%
>{\raggedright\arraybackslash}p{5.0cm} 
>{\centering\arraybackslash}p{1cm}}  
\hline
\textbf{Parameter} & \textbf{Value} \\ \hline
Grid dimensions (x) & 38 \\ \hline
Grid dimensions (y) & 134 \\ \hline
Grid dimensions (z) & 94 \\ \hline
\# of steps (lifespan) & 28 \\ \hline
Max \# of events per step & 8 \\ \hline
\# of individuals & 144 \\ \hline
Tot. \# of genes & 560 \\ \hline
\# of digits per gene & 318 \\ \hline
\# of digits in the genome & 178080 \\ \hline
\# of generations & 32000 \\ \hline
\# of generations per step & 2000 \\ \hline
Crossover probability & 0.5 \\ \hline
(Base) mutation probability & 0.005 \\ \hline
\end{tabular}
\caption{Main system parameters.}
\label{algopars}
\end{minipage}
\hspace{40pt}
\centering
\begin{minipage}[c]{6.0cm} 
\centering
\includegraphics[width=\textwidth]{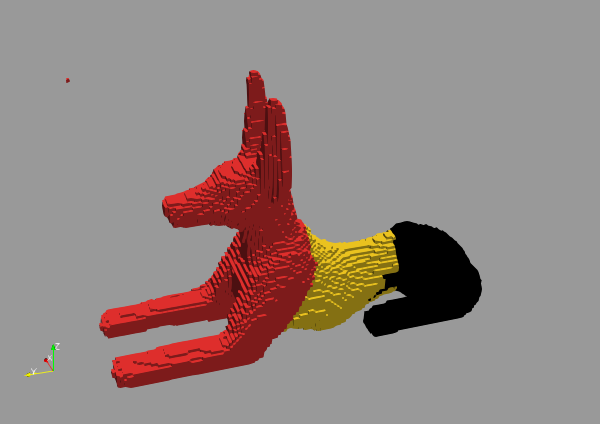} 
\caption{The Belgian Anubis target.}
\label{targetanubis}
\end{minipage}
\end{figure*}

\section{Computer simulations}

\subsection{Experimental setup}

In this section, we outline the computer simulations designed to gather evidence supporting the ESTA hypothesis. These simulations utilise a simplified version of the ET model described in section~\ref{sect:ETmodel}, where only cell proliferation and apoptosis events are considered. During proliferation events, new cells are generated and assigned a colour, which may be thought to represent the cell type. As discussed in section 2, these events are encoded within developmental genes stored in the genome, composed of quaternary digits (0,1,2,3) subject to evolution.

Our approach employs a genetic algorithm to model evolution. In each generation, a population of N=144 individuals with unique genomes undergoes artificial development for a total of N=28 developmental steps, yielding their respective phenotypes. Fitness is evaluated against a predefined target shape: we calculate it by taking the number of correctly coloured cells within the target shape, then subtracting the number of misplaced cells outside the target. This result is divided by the total cell count in the target shape, providing a normalised fitness score. The fitness is used for reproduction, which along with single-point crossover and mutation yield the next generation.

Previous work has demonstrated that the coupling of the ET model and a genetic algorithm gives rise to an evo-devo process able generate large, complex two- and three-dimensional shapes \cite{Fontana12b,Fontana13a}. In this study, we use a bilaterally symmetrical shape modelled after the Egyptian dog-god Anubis, patterned with the colours of the Belgian flag, as our target. This shape, requiring approximately 80000 cells within a 38x134x94 grid, presents a more challenging version of the classic ``French flag'' problem \cite{Wolpert68}.

Each simulation spans 32000 generations. The reproduction step, initially set to 0 at the start of each experiment, increases incrementally by 1 every 2000 generations, reaching a final value of 16 in the last generation. In this context, ``reproduction'' refers to a reference development step used to anchor varying fitness evaluation and mutation profiles, as described below. This increment is externally controlled and uniformly applied to all individuals. Similarly, the genome size starts with 80 genes and is expanded by 30 genes every 2000 generations, culminating in a total of 560 genes by the end of the simulation. Figure~\ref{algopars} provides an overview of the key parameters used in the algorithm for these simulations. The software code is available at: \\
https://github.com/fontalex00/Lxagevolsi/.

\subsection{Genetic algorithm implementation}

\if\afhead1 {\parop{fitness profile}} \fi
The genetic algorithm (GA) can be configured in multiple ways, with three key choices. The first GA choice regards the fitness, which can be evaluated in two ways: either at the point of reproduction or across multiple steps following reproduction. When evaluated across multiple steps, the contribution of each step is weighted by coefficients calculated with the following equation (x is the step, r is the reproduction step):
\[
A(x) =
\begin{cases} 
1 - \frac{1}{1+e^{-0.5(x-r)}} & \text{if } x-r \geq 0, \\
0 & \text{otherwise},
\end{cases}
\]

\noindent These coefficients are and subsequently normalised to ensure that their sum equals 1. This latter option incorporates the concept of kin selection \cite{West07}, an evolutionary strategy that allows individuals to increase their genetic fitness by supporting  the survival of relatives who share common genes. In this model, kin selection is implemented in a straightforward manner: individuals are selected to maintain fitness for a period following reproduction (which in nature may correspond to the period of parental care).

\if\afhead1 {\parop{mutation profile}} \fi
The genetic algorithm consists of wheel selection, crossover, and mutation operators. The crossover is a single-point operation performed with a probability of 0.5. The mutation operator is implemented in two steps. First, the digits to be mutated are randomly selected, with their number determined as a value between 0 and the product of the \textit{base mutation probability} (see Figure~\ref{algopars}) and the total number of genome digits. Next, a random mutation is applied to each selected digit with a second probability $p_2$. This scheme provides computational advantages by increasing variance and accelerating the evolutionary process. The second GA choice regards probability $p_2$, which can either be fixed (= 0.5) or dynamically increased based on a sigmoid function (x is the onset value of the gene to which the digit belongs, r is the reproduction step):
\[
B(x) = \frac{1}{1+e^{-0.5(x-r-2)}}.
\]

\if\afhead1 {\parop{gene onsets}} \fi
The last GA choice regards the gene onsets. These are initalised using the formula: $\text{onset}(c_i) = D(c_i)$, where $D(c_i) = (c_i \mod \text{LIFESPAN})$ represents the onset default value, $c_i$ represents the sequence position of the gene in the genome, and $\text{LIFESPAN}$ denotes the total number of development steps (28 in our simulations). Following this initialisation, the gene onsets can be either allowed to evolve freely or kept fixed to the $D(c_i)$ values.

\begin{table}[h]
\vskip 0.25cm
\centering
\hspace*{-0.0cm}
\small
\begin{tabular}{l
|P{1.2cm} P{1.2cm}
|P{1.2cm} P{1.2cm} 
|P{1.2cm} P{1.2cm}}
\hline
\textbf{Scenario} & 
\multicolumn{2}{|c}{\textbf{BASE}} & 
\multicolumn{2}{|c}{\textbf{VMUT}} & 
\multicolumn{2}{|c}{\textbf{ESTA}} \\ \hline
\textbf{F and M profiles} & 
\textbf{FP} & 
\textbf{MP} & 
\textbf{FP} & 
\textbf{MP} & 
\textbf{FP} & 
\textbf{MP} \\ \hline
\textbf{step r-1} & 0.00 & 0.50 & 0.00 & 0.18 & 0.00 & 0.18 \\ \hline
\textbf{step r-0} & 1.00 & 0.50 & 1.00 & 0.26 & 0.30 & 0.26 \\ \hline
\textbf{step r+1} & 0.00 & 0.50 & 0.00 & 0.37 & 0.23 & 0.37 \\ \hline
\textbf{step r+2} & 0.00 & 0.50 & 0.00 & 0.50 & 0.16 & 0.50 \\ \hline
\textbf{step r+3} & 0.00 & 0.50 & 0.00 & 0.62 & 0.11 & 0.62 \\ \hline
\textbf{step r+4} & 0.00 & 0.50 & 0.00 & 0.73 & 0.07 & 0.73 \\ \hline
\textbf{step r+5} & 0.00 & 0.50 & 0.00 & 0.81 & 0.04 & 0.81 \\ \hline
\textbf{step r+6} & 0.00 & 0.50 & 0.00 & 0.88 & 0.02 & 0.88 \\ \hline
\textbf{Evolvable onsets} & 
\multicolumn{2}{|c}{\textbf{NO}} & 
\multicolumn{2}{|c}{\textbf{NO}} & 
\multicolumn{2}{|c}{\textbf{YES}} \\ \hline
\end{tabular}
\vskip 0.25cm
\caption{
Fitness profiles (FP), mutation profiles (MP) and evolvable onset choice for the three scenarios (r is the reproduction step).}
\label{scentable}
\end{table}

\subsection{Three scenarios}

\if\afhead1 {\parop{scenario BASE}} \fi
The simulations are based on three distinct scenarios, as depicted in Table~\ref{scentable}. In \textbf{Scenario BASE}, fitness is evaluated at a single point (the reproduction step), the mutation rate is constant and the onsets are set to the default values $D(c_i)$. After reproduction, developmental events continue to occur unchecked, with no further impact on fitness. Consequently, the content of these events becomes effectively pseudorandom, leading to mostly detrimental effects on fitness. This setup mirrors the conditions used in our earlier simulations reported in \cite{Fontana14} and discussed further in \cite{Fontana24a}, though the lifespan in the present study has been extended to 28 steps. We expect to see a drastic deterioration in fitness after reproduction.

\if\afhead1 {\parop{scenario VMUT}} \fi
In \textbf{Scenario VMUT}, fitness is evaluated at a single point and the onsets are fixed, as in Scenario BASE, but the mutation rate is onset-dependent. Specifically, the mutation rate for each genome digit depends on the onset of the gene to which the digit belongs, following the profile defined by function B. This design shifts the majority of mutations to the post-reproductive period, reducing the risk of deleterious mutations impairing an individual's fitness before reproduction. While this strategy enhances reproductive success by protecting pre-reproductive fitness, it may come at the cost of reduced evolvability, as fewer mutations occur during the critical phase when beneficial variations could arise and drive adaptive innovation.

\if\afhead1 {\parop{scenario ESTA}} \fi
Finally, in \textbf{Scenario ESTA}, which embodies the evolvable soma hypothesis, fitness is evaluated across multiple steps starting from the reproduction point, capturing the potential influence of kin selection \cite{West07}. The mutation rate in this scenario is variable, with the rate for each genome digit determined by the onset of the gene it belongs to, following the profile defined by function B. As for scenario VMUT, this shifts the majority of mutations to the post-reproductive period. Finally, gene onsets are allowed to evolve. For each scenario, a total of five independent evolutionary runs have been conducted.

\section{Results and discussion}

\begin{figure*}[p] 
\begin{center} \hspace*{-0.0cm}
\includegraphics[width=15.0cm]{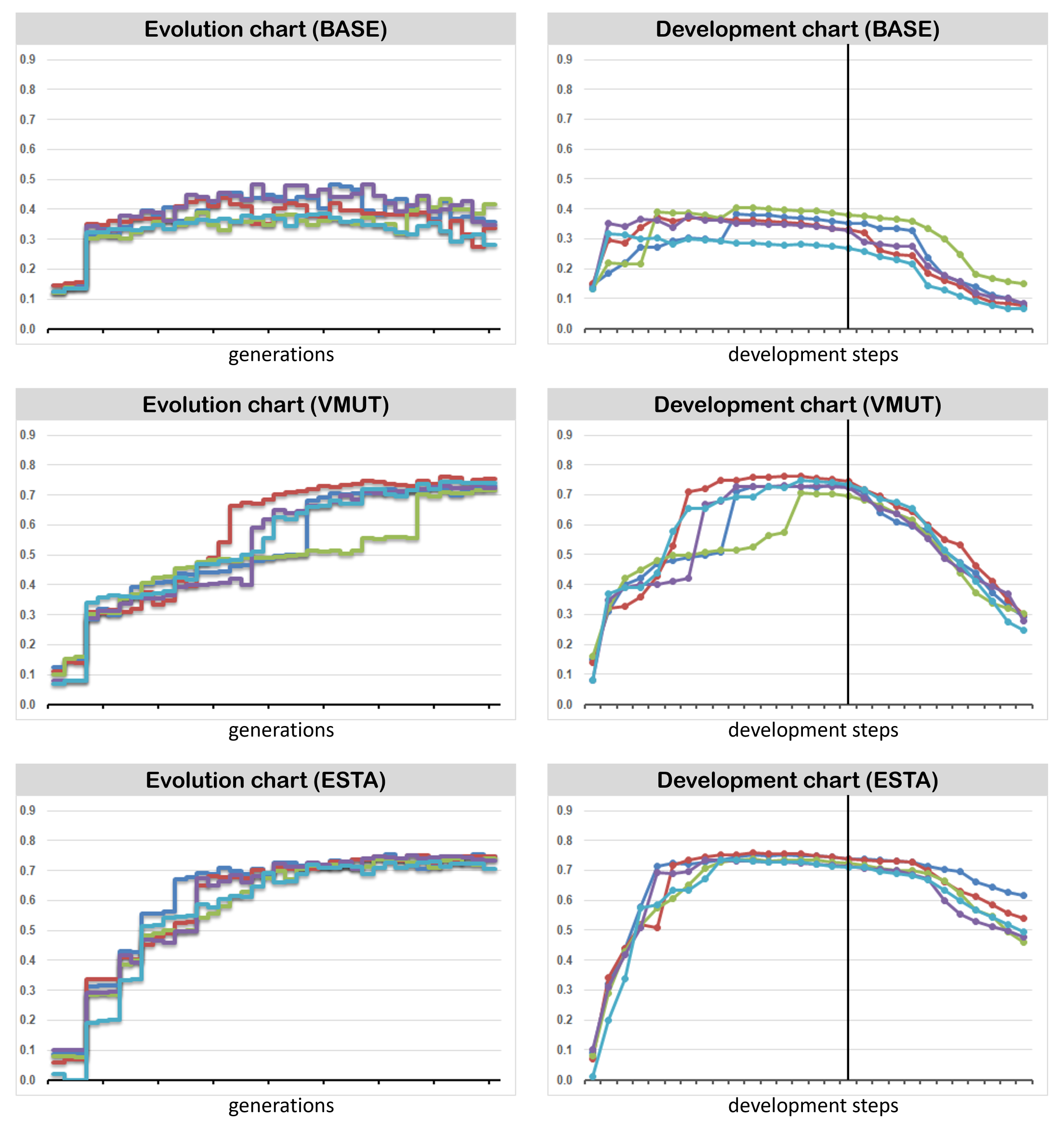}
\caption{
\textbf{Evo-devo charts.} For each scenario, the left panels display the mean reproduction fitness value of the population during evolution, for the five independent evolutionary runs. The right panels show the mean fitness value of the population during development, captured in the last generation, for the five independent evolutionary runs. For all panels, the y-axis represents the fitness value. In the left panels, the x-axis shows the generations, with each of the 8 ticks corresponding to 4000 generations, for a total of 32000 generations. In the right panels, the x-axis represents the development steps, ranging from 0 to 27, with a vertical line indicating the reproduction step at the end of the simulation.
}
\label{evodevocrt}
\end{center} 
\end{figure*}

\begin{figure*}[t] 
\begin{center} \hspace*{-0.0cm}
\includegraphics[width=14.0cm]{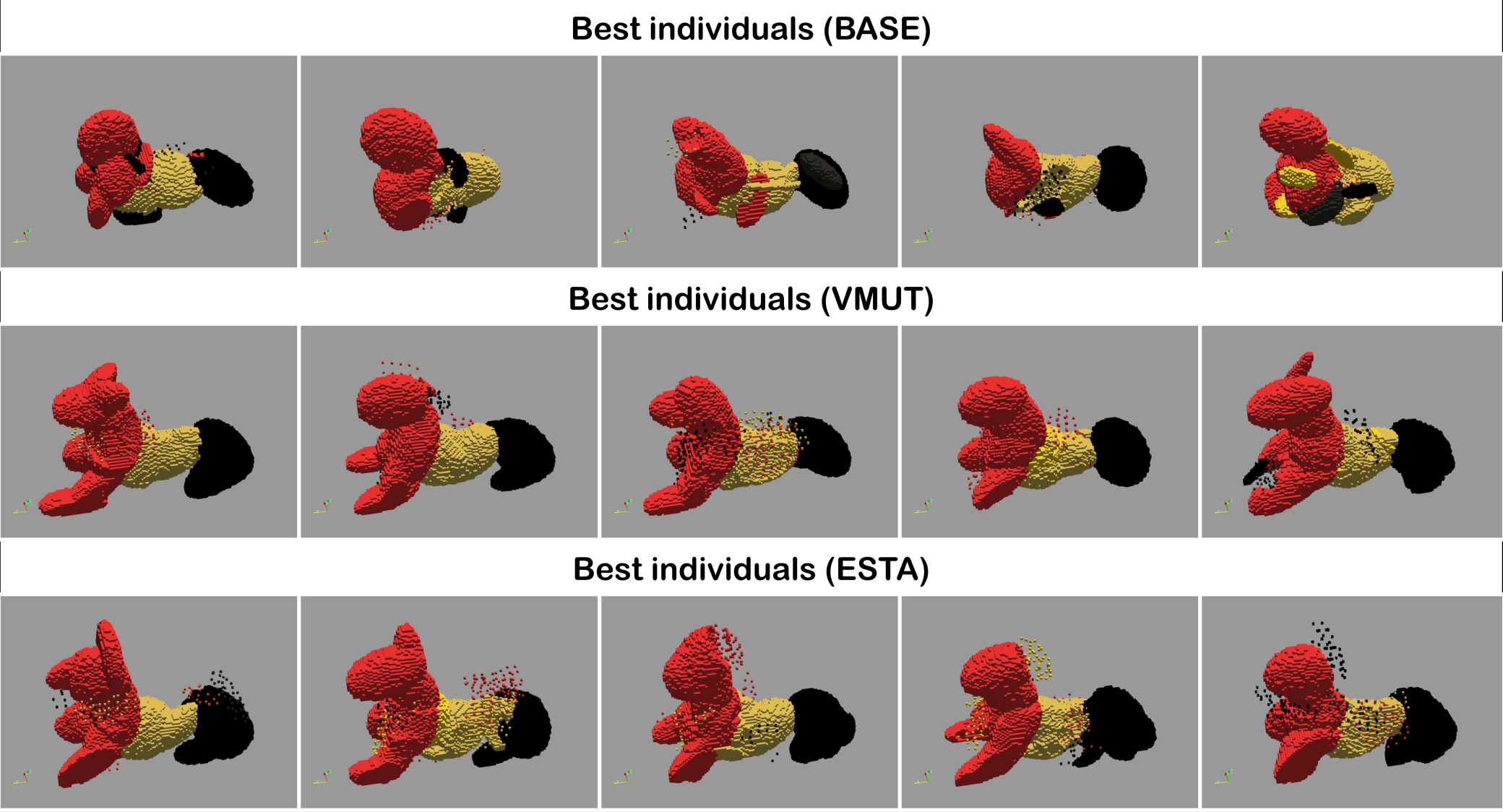}
\caption{
\textbf{Best individuals.} Best individuals evolved under each of the three scenarios, with development shown at the final generation and at the reproduction step, where fitness is close to its peak. The Belgian Anubis target shape (dog with the colours of the Belgian flag), appears in the top-left frame for reference.
}
\label{someguys}
\end{center} 
\end{figure*}

\begin{figure*}[th!] 
\begin{center} \hspace*{-0.0cm}
\includegraphics[width=16.0cm]{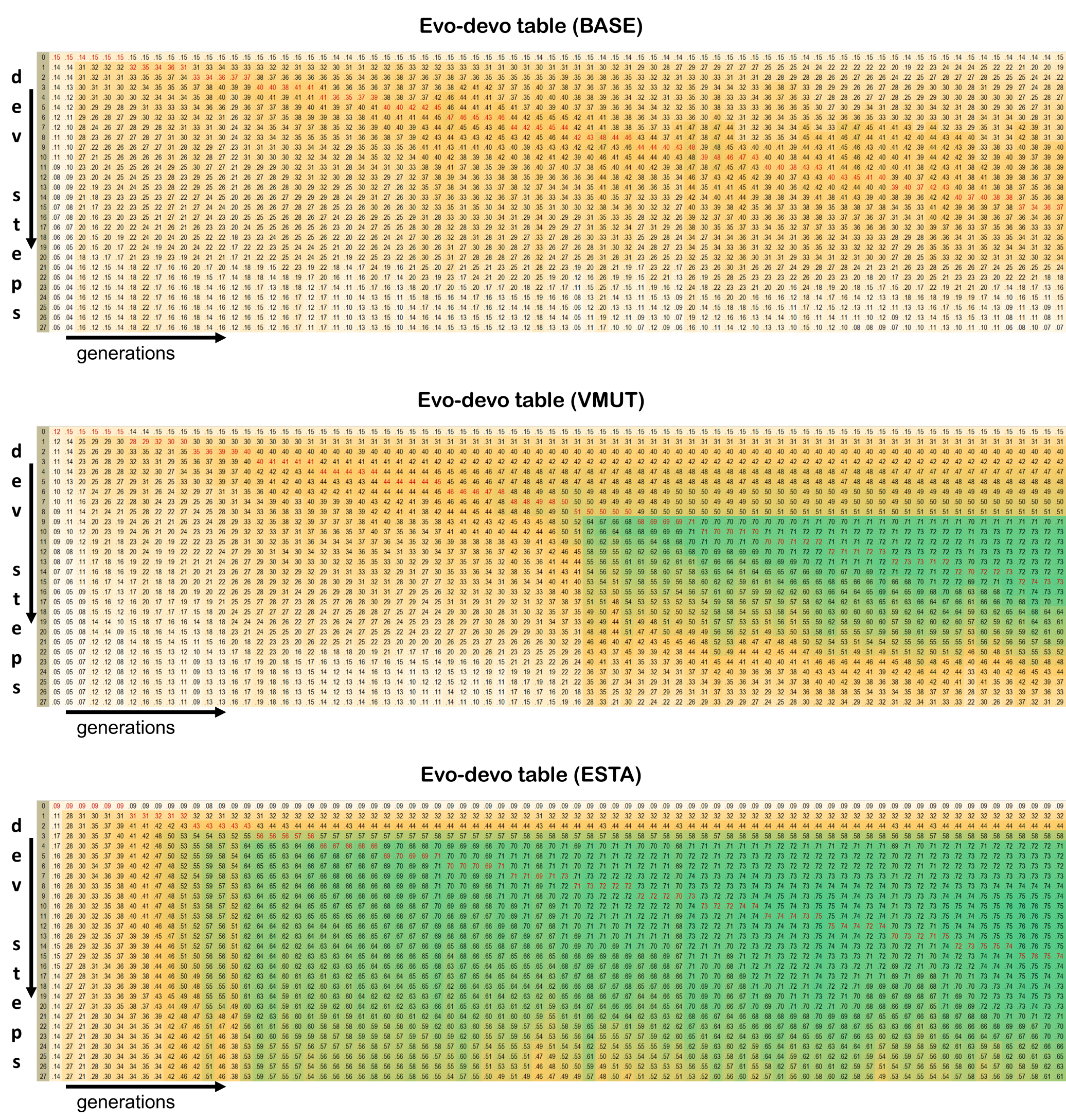}
\caption{
\textbf{Evo-devo tables}. The tables show the evo-devo dynamics of mean population fitness across a single evolutionary run for each of the three scenarios. Red numbers indicate fitness values at each reproduction step. Each table's plot features 80 columns representing generation increments of 400, from 0 to 32000, while rows correspond to the 28 developmental steps.
}
\label{evodevotab}
\end{center} 
\end{figure*}

This section presents and discusses the main experimental results for the three scenarios. Figure~\ref{evodevocrt} presents the evo-devo fitness dynamics, for five independent evolutionary runs, across the three scenarios. Table~\ref{devoprof} presents the mean fitness values of the population alongside the fitness values of the best individuals at various development steps, for the three scenarios. Figure~\ref{someguys} displays the best individuals evolved under each of the three scenarios, with development shown at the final generation and at the reproduction step, where fitness is close to its peak. Figure~\ref{evodevotab} shows the evo-devo dynamics of mean population fitness across a single evolutionary run for each of the three scenarios. 

\begin{table*}[t]
\centering
\hspace*{-0.0cm}
\small
\begin{tabular}{l
|P{1.2cm} P{1.2cm}
|P{1.2cm} P{1.2cm} 
|P{1.2cm} P{1.2cm}}
\hline
\textbf{scenario} & 
\multicolumn{2}{|c}{\textbf{BASE}} & 
\multicolumn{2}{|c}{\textbf{VMUT}} & 
\multicolumn{2}{|c}{\textbf{ESTA}} \\ \hline
\textbf{mean/best} & 
\textbf{mean} & 
\textbf{best} & 
\textbf{mean} & 
\textbf{best} & 
\textbf{mean} & 
\textbf{best} \\ \hline
\textbf{step  8} & 0.34 & 0.46 & 0.63 & 0.63 & 0.74 & 0.75 \\ \hline
\textbf{step 10} & 0.36 & 0.50 & 0.69 & 0.69 & 0.74 & 0.76 \\ \hline
\textbf{step 12} & 0.35 & 0.50 & 0.70 & 0.72 & 0.74 & 0.76 \\ \hline
\textbf{step 14} & 0.35 & 0.49 & 0.73 & 0.75 & 0.73 & 0.77 \\ \hline
\textbf{step 16} & 0.33 & 0.49 & 0.73 & 0.76 & 0.73 & 0.77 \\ \hline
\textbf{step 18} & 0.30 & 0.47 & 0.67 & 0.73 & 0.71 & 0.77 \\ \hline
\textbf{step 20} & 0.28 & 0.45 & 0.62 & 0.71 & 0.70 & 0.77 \\ \hline
\textbf{step 22} & 0.19 & 0.35 & 0.52 & 0.63 & 0.65 & 0.77 \\ \hline
\textbf{step 24} & 0.13 & 0.27 & 0.42 & 0.55 & 0.59 & 0.77 \\ \hline
\textbf{step 26} & 0.10 & 0.25 & 0.33 & 0.46 & 0.54 & 0.77 \\ \hline
\end{tabular}
\vskip 0.20cm
\caption{
\textbf{Development trajectories.} Table showing the mean fitness values of the population alongside the fitness values of the best individuals at various development steps, captured at the last generation, for the three scenarios. Step 16 represents the reproduction step.
}
\label{devoprof}
\end{table*}

\if\afhead1 {\parop{evol /varmut pf /BASE vs ESTA}} \fi
A first key observation from Figure~\ref{evodevocrt}, left panels, is that scenario ESTA achieves higher fitness levels both at reproduction and in later steps compared to scenario BASE. This advantage may be attributed to a ``progressive freezing'' effect in the early developmental steps, as the mutation rate for genes active in these steps approaches zero. Consequently, these early steps become almost frozen, creating a stable foundation for the subsequent steps to build upon. In contrast, if the entire genome remains mutable, changes in early steps can destabilise beneficial adaptations in later steps, hindering their fixation in the genome. This result is consistent with the principle of terminal addition \cite{Gould77}, suggesting that evolutionary innovations tend to occur later in life stages. This effect is also observed in scenario VMUT, which similarly implements a flexible mutation rate.

\if\afhead1 {\parop{evol /varmut popul variance /BASE vs VMUT and ESTA}} \fi
In scenarios ESTA and VMUT the difference between the fitness of the top individuals and the mean population fitness is less pronounced than in scenario BASE (Table~\ref{devoprof}). This outcome can be attributed to the progressive freezing effect discussed above: earlier developmental steps experience few mutations, resulting in reduced variability in the population. Thus, scenarios ESTA and VMUT appear more aligned with modelling natural evolution, where the health and resilience of the entire population are essential. This can be reconducted to the variable, onset-dependent mutation rate, which appears to promote overall stability and collective robustness, ensuring a sustainable population rather than focusing on top performers alone.

\if\afhead1 {\parop{evol /spreval and evolonset /VMUT vs ESTA}} \fi
Fitness progression during evolution is noticeably faster in the ESTA scenario compared to VMUT. This enhanced efficiency in the evolutionary process appears to stem from the combination of a distributed fitness evaluation and evolvable gene onsets. In~\cite{Fontana24a}, we proposed that once genes are ``tested'' by evolution during the ageing phase, their associated onset values could be adjusted, shifting their expression to earlier steps where they would exert a greater influence on fitness. Our experimental results suggest that this anticipated effect is taking place. Nonetheless, further testing is required to validate this finding.

\if\afhead1 {\parop{evol /spreval fitness drops}} \fi
In scenarios BASE and VMUT some drops in fitness value (Figure~\ref{evodevocrt}) occur when the reproduction step is advanced. Let's denote the evaluation step before a change as T. Prior to this change, genes with onset values greater than T were not selected for by evolution, leaving these regions effectively pseudorandom. When T increases to T+1, these previously unselected genes now influence the phenotype, often detrimentally, leading to a fitness drop. This phenomenon is absent in scenario ESTA, possibly because genes influencing later developmental steps remain subject to evolutionary pressure, albeit reduced. Consequently, these genes are already partially optimised, decreasing the likelihood that their incorporation will cause substantial fitness declines. This further supports the notion that ageing functions as a ``trial'' period, offering a lower-risk setting for evolutionary experimentation.

\if\afhead1 {\parop{devo /spreval post-repro decline /BASE and VMUT vs ESTA}} \fi
As illustrated in the right panels of Figure~\ref{evodevocrt} and in Figure\ref{evodevotab}, both the BASE and VMUT scenarios exhibit significant fitness deterioration following reproduction, as genes expressed during this period are no longer subject to evolutionary pressure. In contrast, the ESTA scenario also shows a notable decline in post-reproductive fitness due to a heightened mutation rate; however, this decline is milder and more controlled, reflecting the fitness evaluation profile of the ESTA model. As intended, this captures the idea that individuals must remain fit as long as they have parental care responsibilities, effectively extending their fitness beyond reproduction for a defined period.


\if\afhead1 {\parop{devo /spreval super-agers /ESTA}} \fi
One interesting observation concerns the best individuals in the population under the ESTA scenario (Table~\ref{devoprof}). These individuals exhibit limited ageing, with their fitness remaining constant or experiencing only mild decline after reproduction. This minority of individuals could be termed ``super-agers'' \cite{deGodoy21}, in analogy with the similar phenomenon observed in humans. However, the comparison is somewhat loose, as human super-agers still experience fitness decline and eventually die, whereas our model is far simpler and lacks the complexity of real organisms. While this trait may provide advantages to the individual, it could be detrimental to the population as a whole, assuming that evolution depends on variation introduced through ageing and that the absence of ageing would halt evolution. 

\section{Conclusions and future work}

In this study, we present evidence supporting the Evolvable Soma Theory of Ageing through computer simulations. We conducted experiments on a platform where genes are associated with an onset value that dictates when they are expressed. Three scenarios have been tested: one with single-point fitness evaluation, constant mutation rate and fixed gene onsets; one with single-point fitness evaluation, onset-dependent mutation rate and fixed gene onsets; and one with spread fitness evaluation, onset-dependent mutation rate and evolvable gene onsets. The last scenario, which embodies the evolvable soma hypothesis, demonstrates superior performance in both algorithmic efficiency and biological plausibility compared to the others.

In this work, when the mutation rate was dependent on the onset value of developmental genes, the mutation profiles were fixed and hard-wired. A potential direction for future research is to encode in the genome the mutation rate for each developmental gene, allowing it to evolve over time. This approach could lead to the emergence of mutation profiles similar to those used in the present study---specifically, a gene onset-dependent mutation rate with the probability mass shifted toward post-reproductive steps.

In~\cite{Fontana24a}, we proposed that once genes are effectively ``tested'' by evolution during ageing, evolution could adjust their associated onset values, shifting their expression to earlier steps where they would have a stronger impact on fitness. Although our preliminary findings suggest support for this hypothesis, confirming it will require monitoring gene dynamics over multiple generations to detect any potential shifts in timing and expression. This line of investigation, which could shed light on how evolutionary processes optimise gene expression timing, will be explored in future work.

\bibliographystyle{plain}
\bibliography{lxagevolsi}
 
\end{document}